\documentstyle[aps,preprint,psfig,floats]{revtex}
\newcommand{\MET}{\hbox{$\rlap{\,/}E_T $}}
\begin{document}
\preprint{FERMILAB-Conf-95/250-E}
\title{Search for Anomalous $WW$ and $WZ$ production at {D\O}\thanks{Submitted
to the International
Europhysics Conference on High Energy Physics (HEP 95), Brussels, Belgium,
 27 July - 2 August, 1995, and
the 17th International Symposium on Lepton-Photon Interactions, Beijing, China,
August 10 - 15, 1995.}}

\author{D\O\ Collaboration}

\date{\today}

\maketitle

\begin{abstract}
We present a preliminary result from a search for anomalous
$WW$ and $WZ$ production in $p\bar{p}$ collisions at $\sqrt{s}=1.8$ TeV
using $p\bar{p}\rightarrow e\nu jj$ events
observed during the 1992--1993 run of the Fermilab Tevatron collider.
A fit to the $p_T$ spectrum of $W(e\nu)$ yields
direct limits on the CP--conserving
anomalous $WW\gamma$ and $WWZ$ coupling parameters of
$-0.89<\Delta\kappa<1.07~~(\lambda=0)$ and
$-0.66<\lambda<0.67~~(\Delta\kappa=0)$
at the $95\ \%$ confidence level,
assuming that the $WWZ$ coupling parameters are equal
to the $WW\gamma$ coupling parameters,
and a form factor scale $\Lambda=1.5$~TeV.
\end{abstract}

\vskip 0.5in
\noindent
\begin{center}
\small
S.~Abachi,$^{12}$
B.~Abbott,$^{34}$
M.~Abolins,$^{23}$
B.S.~Acharya,$^{41}$
I.~Adam,$^{10}$
D.L.~Adams,$^{35}$
M.~Adams,$^{15}$
S.~Ahn,$^{12}$
H.~Aihara,$^{20}$
J.~Alitti,$^{37}$
% ...
\end{center}

\vskip 0.50cm
\small
\centerline{$^{1}$Universidad de los Andes, Bogot\'{a}, Colombia}
\centerline{$^{2}$University of Arizona, Tucson, Arizona 85721}
\centerline{$^{3}$Brookhaven National Laboratory, Upton, New York 11973}
\centerline{$^{4}$Brown University, Providence, Rhode Island 02912}
\centerline{$^{5}$University of California, Davis, California 95616}
\centerline{$^{6}$University of California, Irvine, California 92717}
\centerline{$^{7}$University of California, Riverside, California 92521}
\centerline{$^{8}$LAFEX, Centro Brasileiro de Pesquisas F{\'\i}sicas,
                  Rio de Janeiro, Brazil}
\centerline{$^{9}$CINVESTAV, Mexico City, Mexico}
\centerline{$^{10}$Columbia University, New York, New York 10027}
\centerline{$^{11}$Delhi University, Delhi, India 110007}
\centerline{$^{12}$Fermi National Accelerator Laboratory, Batavia,
                   Illinois 60510}
\centerline{$^{13}$Florida State University, Tallahassee, Florida 32306}
\centerline{$^{14}$University of Hawaii, Honolulu, Hawaii 96822}
\centerline{$^{15}$University of Illinois at Chicago, Chicago, Illinois 60607}
\centerline{$^{16}$Indiana University, Bloomington, Indiana 47405}
\centerline{$^{17}$Iowa State University, Ames, Iowa 50011}
\centerline{$^{18}$Korea University, Seoul, Korea}
\centerline{$^{19}$Kyungsung University, Pusan, Korea}
\centerline{$^{20}$Lawrence Berkeley Laboratory and University of California,
                   Berkeley, California 94720}
\centerline{$^{21}$University of Maryland, College Park, Maryland 20742}
\centerline{$^{22}$University of Michigan, Ann Arbor, Michigan 48109}
\centerline{$^{23}$Michigan State University, East Lansing, Michigan 48824}
\centerline{$^{24}$Moscow State University, Moscow, Russia}
\centerline{$^{25}$University of Nebraska, Lincoln, Nebraska 68588}
\centerline{$^{26}$New York University, New York, New York 10003}
\centerline{$^{27}$Northeastern University, Boston, Massachusetts 02115}
\centerline{$^{28}$Northern Illinois University, DeKalb, Illinois 60115}
\centerline{$^{29}$Northwestern University, Evanston, Illinois 60208}
\centerline{$^{30}$University of Notre Dame, Notre Dame, Indiana 46556}
\centerline{$^{31}$University of Oklahoma, Norman, Oklahoma 73019}
\centerline{$^{32}$University of Panjab, Chandigarh 16-00-14, India}
\centerline{$^{33}$Institute for High Energy Physics, 142-284 Protvino, Russia}
\centerline{$^{34}$Purdue University, West Lafayette, Indiana 47907}
\centerline{$^{35}$Rice University, Houston, Texas 77251}
\centerline{$^{36}$University of Rochester, Rochester, New York 14627}
\centerline{$^{37}$CEA, DAPNIA/Service de Physique des Particules, CE-SACLAY,
                   France}
\centerline{$^{38}$Seoul National University, Seoul, Korea}
\centerline{$^{39}$State University of New York, Stony Brook, New York 11794}
\centerline{$^{40}$SSC Laboratory, Dallas, Texas 75237}
\centerline{$^{41}$Tata Institute of Fundamental Research,
                   Colaba, Bombay 400005, India}
\centerline{$^{42}$University of Texas, Arlington, Texas 76019}
\centerline{$^{43}$Texas A\&M University, College Station, Texas 77843}

\normalsize
%end

\bigskip

\begin{figure}
\centerline{\psfig{file=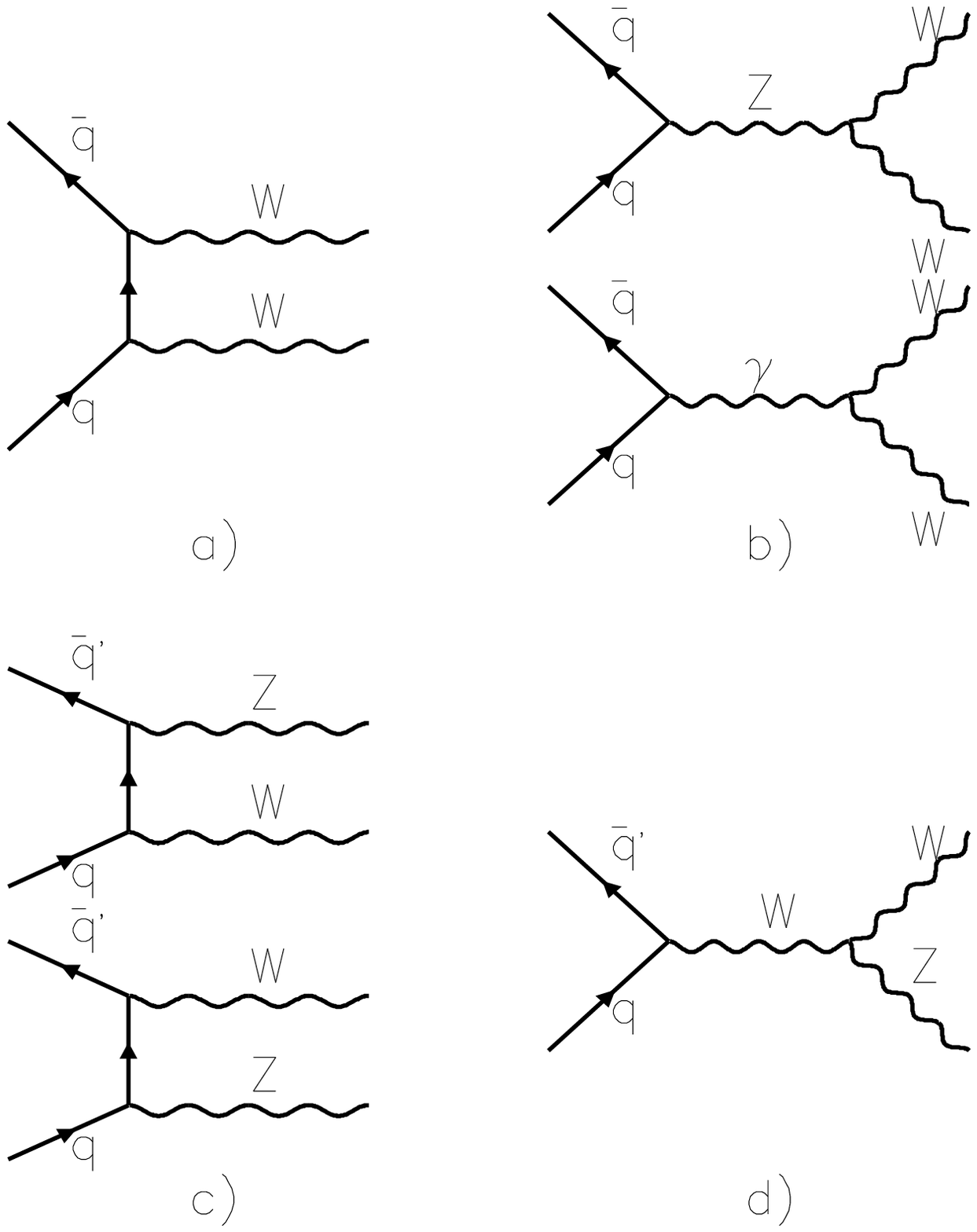,bbllx=0pt,bblly=100pt,bburx=600pt,bbury=750pt,
width=15cm}}
\caption{Leading order Feynman diagrams for $q\bar{q}\rightarrow WW$ (a,b) and
$q\bar{q^\prime}\rightarrow WZ$ (c,d)}
\label{Feynman}
\end{figure}

\begin{figure}
\centerline{\psfig{file=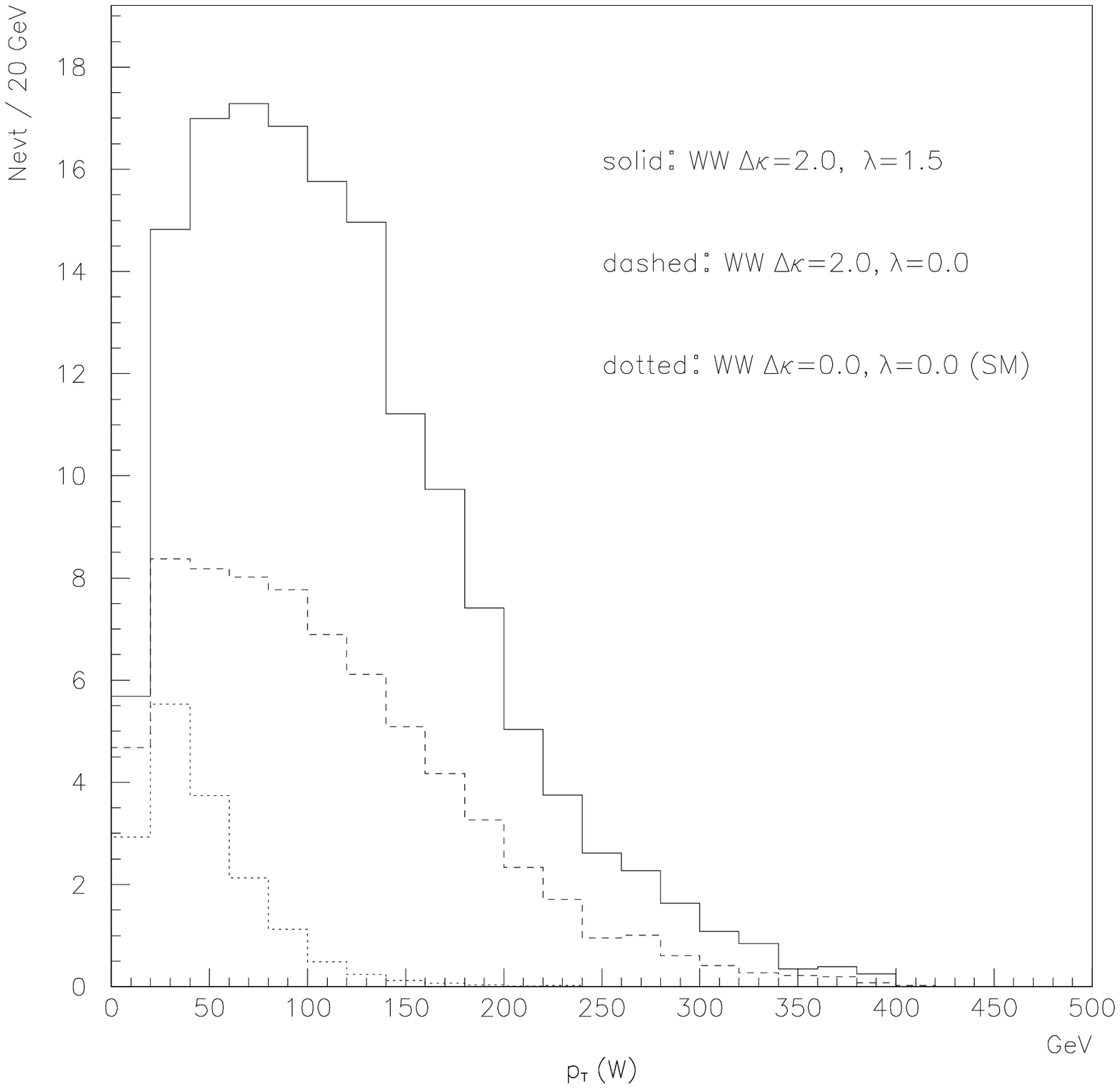,bbllx=0pt,bblly=140pt,bburx=550pt,bbury=670pt,
width=15cm}}
\caption{$p_T$ distributions of Monte Carlo $WW\rightarrow e\nu jj$
events
with various coupling parameters.
The dotted line represents the Standard Model (SM) couplings.
The cross section increases and the $p_T$
spectrum becomes harder with anomalous coupling parameters.
The samples are
normalized to $13.7\ {\rm pb^{-1}}.$}
\label{PTW_MC}
\end{figure}

The self-interaction of electroweak gauge bosons is
a direct consequence of the
non-Abelian gauge theory of the Standard Model (SM) and can be tested
through study of gauge boson pair ($W\gamma$, $Z\gamma$, $WW$ and $WZ$)
production
in $p\bar{p}$ collisions at $\sqrt{s}=1.8$ TeV~\cite{DPF}.
The self-interaction coupling parameters are given precisely in the SM.
Any deviation
of the parameters from the SM values signals physics beyond the SM.
Figure~\ref{Feynman} shows leading order Feynman diagrams
of $q\bar{q}\rightarrow WW$ and $q\bar{q}^\prime\rightarrow WZ$ processes.
The $WW$ production process depends strongly on the
$WW\gamma$ and $WWZ$ coupling parameters due to destructive interference
between contributing amplitudes.
This interference prevents the SM $WW$ cross section from violating
unitarity at high energies.
The SM predicts the production cross sections for $p\bar{p}\rightarrow W^+W^-$
and $p\bar{p}\rightarrow W^\pm Z$ at $\sqrt{s}=1.8$ TeV to be 8.4~pb and
2.5~pb,
respectively~\cite{Hagiwara}.
Based on a formalism developed by Hagiwara {\it et.~al}~\cite{Peccei}
the $WW\gamma$ and $WWZ$ interactions beyond the SM can be parametrized by
four independent dimensionless coupling parameters\footnote{In this paper
we only consider CP--conserving couplings.},
$\Delta\kappa_\gamma$ and $\lambda_\gamma$ for the $WW\gamma$ vertex and
$\Delta\kappa_Z$ and $\lambda_Z$ for the $WWZ$ vertex.
For the SM, $\Delta\kappa_\gamma=\lambda_\gamma=\Delta\kappa_Z=\lambda_Z=0.$
Non-zero coupling parameters result in a dramatic
increase of the production
cross section and an enhancement in the transverse momentum
($p_T^W$)  spectrum  of the $W$ boson in the high $p_T$ region
as shown in Fig.~\ref{PTW_MC}.
Thus, a study of the $p_T^W$ spectrum of  $WW$ production leads to
a sensitive test of the $WW\gamma$ and $WWZ$ couplings.
Similarly, the $p_T^W$ spectrum of $WZ$ production provides a direct test
of the $WWZ$ coupling.

The D{\O} collaboration has previously reported
limits on anomalous trilinear gauge boson couplings from three processes
using the data from the 1992--93 Tevatron collider run:
the $WW\gamma$ coupling based on a measurement of $W\gamma$
production~\cite{D0:Wgamma},
$WWZ$ and $WW\gamma$ couplings from a
search for  $W$ boson pair production in dilepton decay modes~\cite{D0:WW},
and
$ZZ\gamma$ and $Z\gamma\gamma$ couplings from
a measurement of $Z\gamma$ production~\cite{D0:Zgamma}.
In this report
we present a new, independent determination of  limits on the anomalous
$WW\gamma$ and $WWZ$
couplings obtained
from a search for $p\bar{p}\rightarrow WW+X$ followed by $W\rightarrow
e\nu$ and $W\rightarrow jj$, where $j$ represents a jet,
and $p\bar{p}\rightarrow WZ+X$ followed by
$W\rightarrow e\nu$ and $Z\rightarrow jj,$
using the data from the 1992--1993 run,
corresponding to an integrated luminosity of $13.7\pm 0.7\ {\rm pb}^{-1}.$
In this decay mode, $WZ$ events are indistinguishable from
$WW$ events.\footnote{
The SM predicts
$\sigma\cdot B(p\bar{p}\rightarrow W^+W^-\rightarrow
e^\pm\nu jj)=1.23~{\rm pb}$ and
$\sigma\cdot B(p\bar{p}\rightarrow W^\pm
Z\rightarrow e^\pm\nu jj)=0.19~{\rm pb}.$}
The CDF collaboration has reported a similar measurement~\cite{CDF:WW}.

The $WW,WZ\rightarrow e{\nu}jj$ candidates were selected by searching for
events containing a $W\rightarrow e\nu$ decay and two  jets
consistent with $W\rightarrow jj$ or $Z\rightarrow jj.$
The data sample was obtained with a single electron trigger:
an isolated electromagnetic (EM) cluster
with transverse energy $E_{T}^{e}>20~{\rm{GeV}}$.
This EM cluster was required to be within the fiducial region of
the calorimeter
$|\eta|\leq 1.1$ in the central calorimeter, or $1.5\leq |\eta|\leq 2.5$
in the end calorimeters.
Here $\eta$ is the pseudorapidity defined as $\eta=-\ln (\tan(\theta/2)),$
$\theta$ being the polar angle with respect to the beam axis.
The electron cluster had to have
(i) a ratio of EM energy to the total shower energy
greater than 0.9;
(ii) lateral and longitudinal shower shape consistent
with an electron shower;
(iii) the isolation variable of the cluster
less than 0.1, where isolation is
defined as $I  = (E(0.4) - EM(0.2))/EM(0.2),$ and
$E(0.4)$ is the total calorimeter energy inside
a cone of radius
${\cal R}\equiv\sqrt{(\Delta\eta)^2+(\Delta\phi)^2}=0.4$, and
$EM(0.2)$ is the EM energy
inside a cone of 0.2; and
(iv) a matching track in the drift chambers.
The $W\rightarrow e\nu$ decay was identified by
an isolated electron
with $E_{T}^{e}>25~{\rm{GeV}}$ and missing transverse
energy $\MET>25~{\rm{GeV}}$ forming
a transverse mass $M_{T}^{e\nu}>40~{\rm{GeV/c^2}}.$

Jets were reconstructed by applying a cone algorithm with
a radius ${\cal R} = 0.3$ to the calorimeter hits.
This small cone size minimized the probability for two jets from the $W(Z)$
boson  to merge into one cluster in the calorimeter, in particular,
in the high $p_T$ region.
The jets were required to be within  $|\eta|<2.5$
and energy corrections including that
for out-of-cone gluon radiation were applied~\cite{TOP_PRD}.
We required that a candidate event contain at least two jets with
$E_T^j>20$~GeV and that
dijet invariant mass (the largest invariant mass if more than two jets
with $E_T^j>20$~GeV in the event)
satisfy $50<m_{jj}<110~{\rm{GeV/c^{2}}},$
consistent with $W$ and $Z$ masses.
The above selection criteria  yielded 84 candidate events.

The background estimate, summarized in Table~1, includes
contributions from: QCD production of $W+\geq 2j$;
QCD multijet events, where a jet was misidentified as an electron;
$t\bar{t}\rightarrow W^+W^-b\bar{b}\rightarrow e\nu jj X$;
$WW$ with $W\rightarrow \tau\nu$ followed by $\tau\rightarrow e\nu\bar{\nu};$
and $ZX\rightarrow eeX$, where one electron was lost.
The multijet background was estimated from the data by measuring
the \MET\ distribution of a background-dominated sample,
obtained by selecting events containing
an EM cluster which failed at least one of the electron
quality requirements (isolation, shower shape and track-match).
We extrapolated this \MET\ distribution
into the signal region ($\MET>25~{\rm GeV}$)
by normalizing the number of events in the background sample to that in the
candidate sample (without the \MET\ requirement imposed)
in the region of small  \MET $\ (0<\MET<15~{\rm{GeV}})$.
We measured the total number of multijet background
events to be $12.2\pm 2.6.$
The $W+\geq 2j$ background was estimated using
the VECBOS\cite{VECBOS} Monte Carlo
followed by parton fragmentation using the ISAJET\cite{ISAJET} program
and a full detector
simulation based on the GEANT program~\cite{GEANT}.
Using
the dijet invariant mass distributions of
the VECBOS sample and the observed $Wjj$ sample after subtracting the
contribution from the multijet events,
we normalized the number of VECBOS $W+\geq 2j$ events
to the number of observed $Wjj$ events
outside of the signal region $50<m_{jj}<110~{\rm{GeV/c^{2}}}.$
This yielded
the total number of $W+\geq 2j$ background
events (in the signal region) as $62.2\pm 13.0$,
where the uncertainty was due to the normalization ($16\%$)
and the limited
statistics of the Monte Carlo events ($13\%$).
As a cross check of the normalization, we also calculated this background
using the VECBOS prediction for
the $W+\geq 2j$ inclusive cross section and
obtained a consistent result.

The backgrounds due to
$t{\bar t}\rightarrow W^+W^-b{\bar b}$,
$WW\rightarrow \tau\nu jj$ and $ZX\rightarrow eeX$ were
estimated using the ISAJET program followed by the GEANT
detector simulation and found to be small.
The total number of background events was estimated to be $75.5\pm 13.3.$
Thus we observed no statistically significant signal above the background.

\begin{table}
\caption{Summary of $e\nu jj$ data and backgrounds.}
\label{table1}
\begin{tabular}{llcr}
 & & $e\nu jj$ events & \\
\tableline
 & Background source: & &\\
 &$~~~W+\geq 2j$ & $62.2\pm 13.0$ & \\
 &$~~~{\rm multijets}$ & $12.2\pm 2.6$ & \\
 &$~~~t\bar{t}(m_t=180{\ \rm GeV/c^2})$ & $0.87\pm 0.01$  & \\
 &$~~~WW\rightarrow \tau\nu jj$ & $0.19\pm 0.01$ & \\
 &$~~~ZX\rightarrow eeX$ & $0.00^{+0.34}_{-0.00}$ & \\
 &Total Background & $75.5\pm 13.3$ & \\
\tableline
%% & & & \\
 & Data & 84 & \\
\tableline
& SM $WW+WZ$ prediction & $2.9\pm 0.5$ & \\
\end{tabular}
\end{table}

\begin{figure}
\centerline{\psfig{file=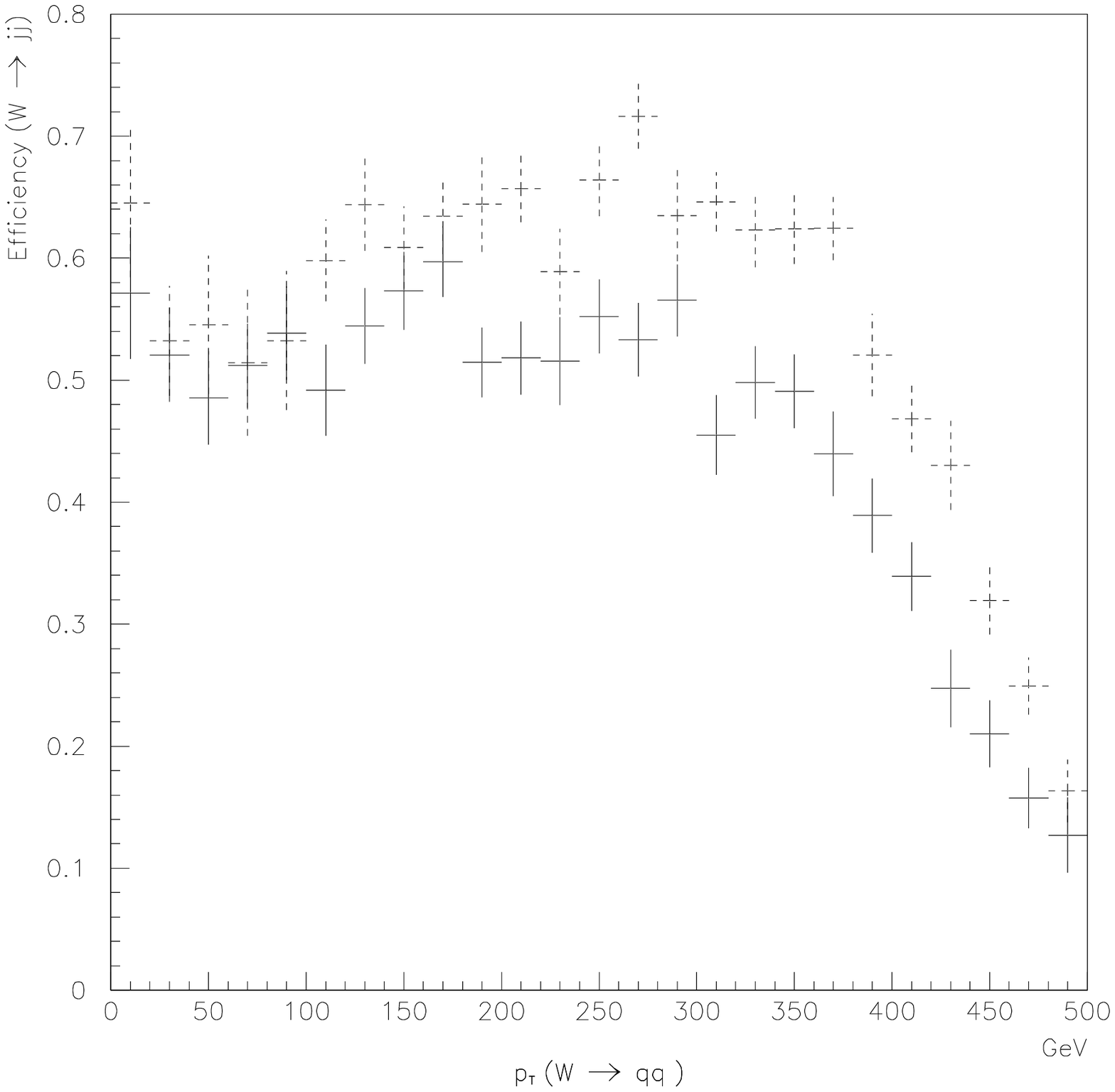,bbllx=0pt,bblly=100pt,bburx=550pt,bbury=700pt,
width=15cm}}
\caption{Total efficiency for $W\rightarrow jj$ selection as a function
of $p_T^W,$ estimated using the ISAJET(solid) and the PYTHIA(dashed)
generators followed by a full detector simulation.}
\label{JET_EFF}
\end{figure}

\begin{figure}
\centerline{\psfig{file=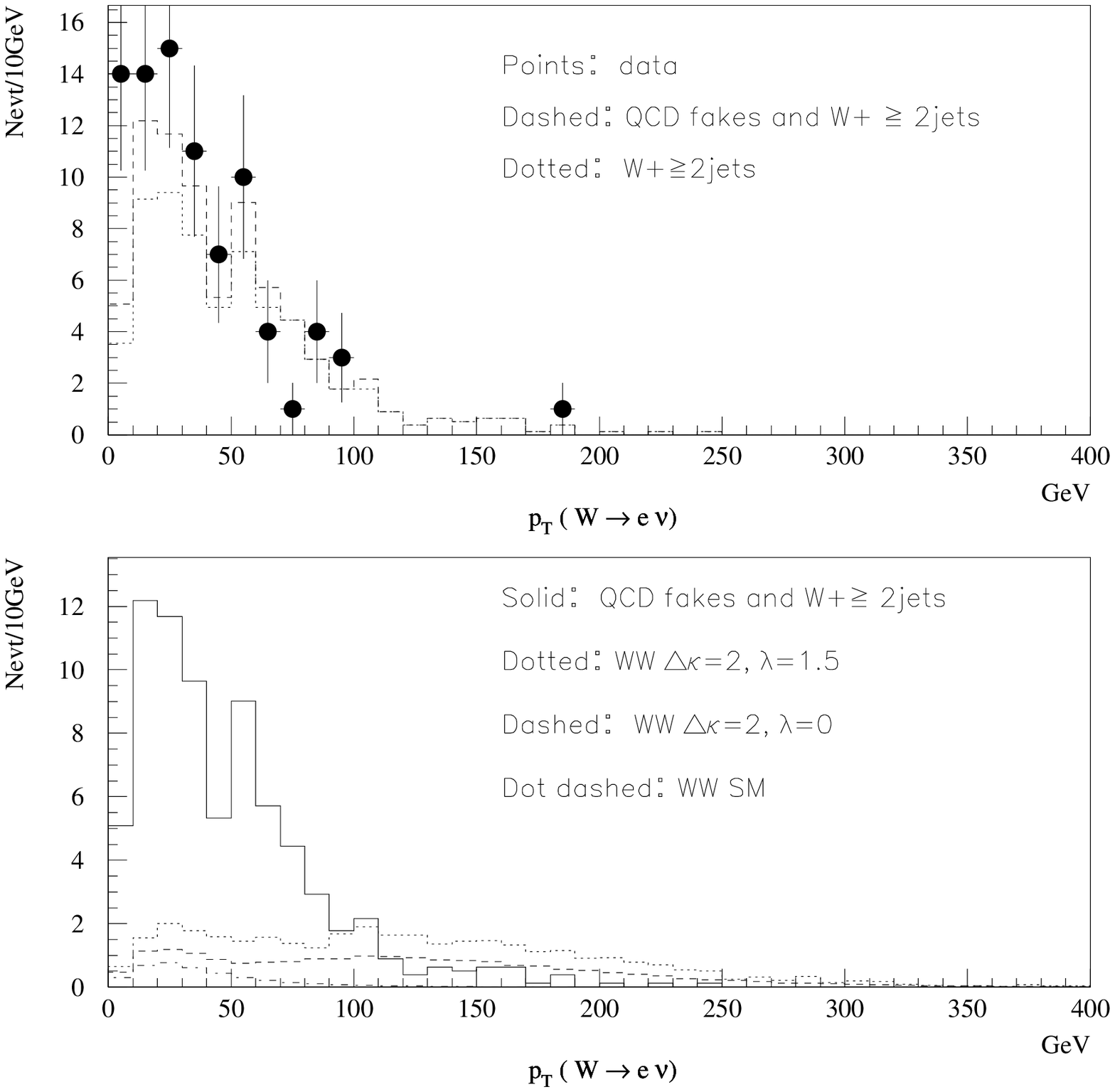,bbllx=0pt,bblly=100pt,bburx=550pt,bbury=700pt,
width=15cm}}
 \caption{ $p_T$ distributions of the $e\nu$ systems. The solid circle
  indicates the observed spectrum. The dashed and dotted lines are background
  estimates from the QCD multi-jet events and $W + \geq 2j$ events, and
  $W + \geq 2j$ events only respectively (top plot).
  The Monte Carlo predictions of $p_T$ spectrum of the $e\nu$ system
  for the SM and non-SM productions are shown in the bottom plot.}
 \label{PTW_DATA}
\end{figure}

The trigger and electron selection efficiencies~\cite{Xsection}
were estimated using
$Z\rightarrow ee$ events.
The jet finding efficiency is a function of $p_T^W$,
due to the $E_T^j$ requirement in the low $p_T^W$ region
and due to the probability for two jets to merge into one in the high
$p_T^W$ region.
Using the ISAJET and PYTHIA~\cite{PYTHIA} event generators
followed by a full detector simulation,
we estimated the efficiency for
$W\rightarrow jj$ selection, including the jet finding efficiency and
the efficiency for the dijet mass requirement,
as a function of $p_T^W,$ shown in Fig.~\ref{JET_EFF}.
In estimating the sensitivity to the anomalous $WW\gamma$ and $WWZ$ coupling
parameters, we used the $W\rightarrow jj$
efficiency obtained from ISAJET, which is
smaller than that from  PYTHIA and therefore gives a conservative
estimate.
We included the difference between the ISAJET and  PYTHIA numbers
in the systematic uncertainty.
We calculated the overall event selection efficiency as a function of
the coupling parameters using the efficiencies described  above
and the  $WW$, $WZ$ Monte Carlo program of
Zeppenfeld~[2,14],
%%\cite{Hagiwara}~\cite{MC:Zeppenfeld},
in which the
processes were generated to leading order, and higher order QCD effects
were approximated by a K-factor of $1+\frac{8}{9}\pi\alpha_s=1.34$.
A dipole form factor with a scale $\Lambda = 1.5~{\rm TeV}$ was
used in the Monte Carlo event generation
(e.g. $\Delta\kappa_\gamma(\hat{s})=\Delta\kappa/(1+\hat{s}/\Lambda^2)^2$,
where
$\hat{s}$ is the square of the invariant mass of the $WW$ or $WZ$ system).
We simulated the $p_T$ distribution of the $WW$ and $WZ$ systems
using the observed $p_T^Z$ spectrum  in our inclusive
$Z\rightarrow ee$ data sample.
We calculated the total efficiency with the SM couplings to be
$0.15\pm 0.02$ for $WW$ and
$0.16\pm 0.02$ for $WZ$.
Thus the total number of expected SM events
was  $2.9\pm 0.5$:
$2.5\pm 0.5$ for $WW$ and $0.4\pm 0.1$ $WZ$.
Using these efficiencies and the background-subtracted signal,
we set the upper limit on the cross section times branching
fraction of
$\sigma B(W^+W^-\rightarrow e^\pm \nu jj) + \sigma B(W^\pm
Z\rightarrow e^\pm \nu jj)$
for the SM couplings to be 17~pb
at the $95\%$ confidence level (CL).
Figure~\ref{PTW_DATA} shows
the $p_T$ distribution of the $e\nu$ system.

\begin{figure}
\centerline{\psfig{file=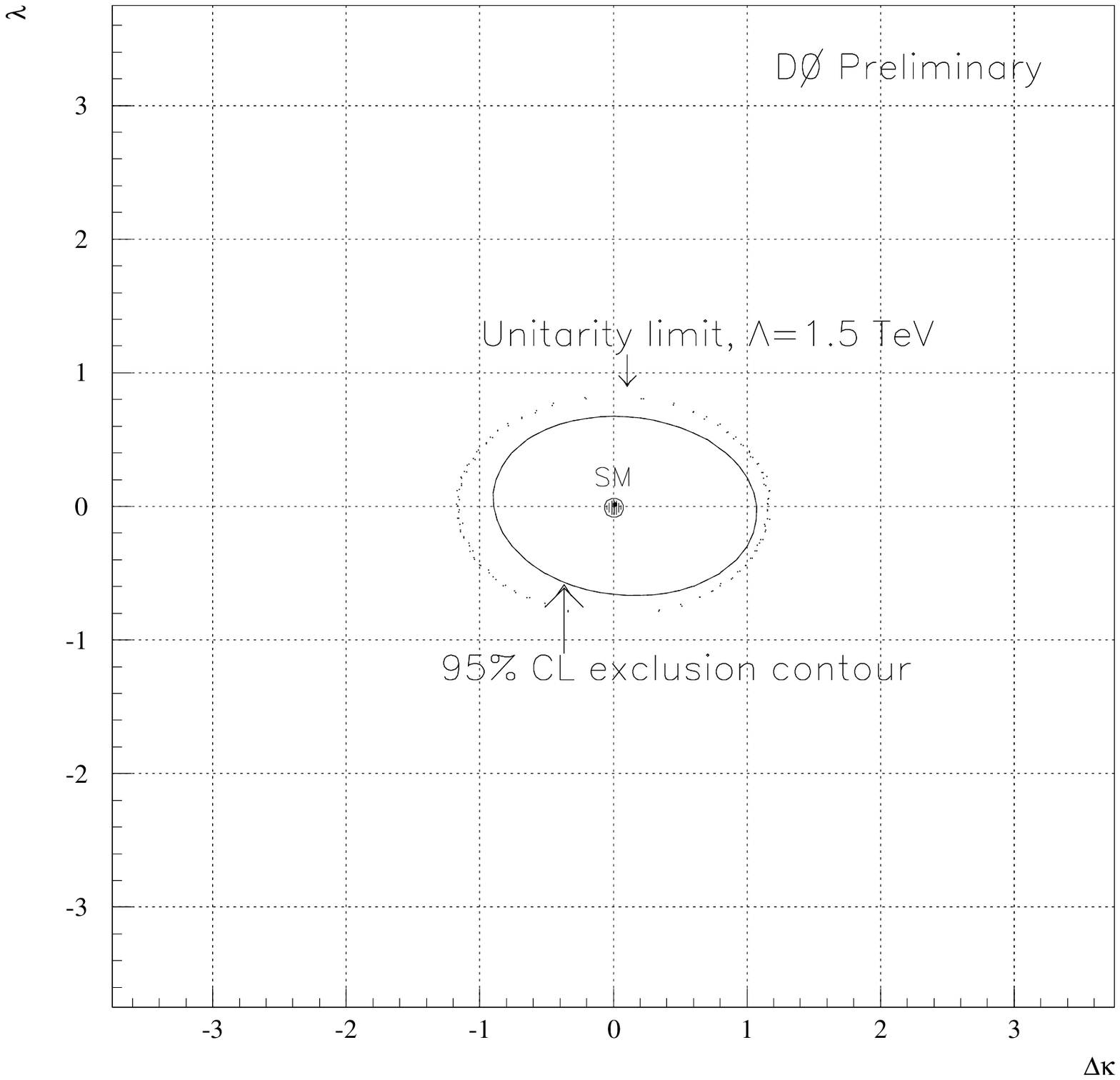,bbllx=0pt,bblly=100pt,bburx=550pt,bbury=700pt,
width=15cm}}
\caption{Limit contour (solid line)  on CP--conserving anomalous coupling
parameters at the $95\%$ CL, assuming
$\Delta\kappa\equiv\Delta\kappa_{\gamma}=\Delta\kappa_{Z}$ and
$\lambda\equiv\lambda_{\gamma}=\lambda_{Z}$.
The constraint imposed by the S-matrix unitarity for
$\Lambda=1.5$ TeV is also shown (dotted line).}
\label{WW:contour}
\end{figure}

\begin{figure}
\centerline{\psfig{file=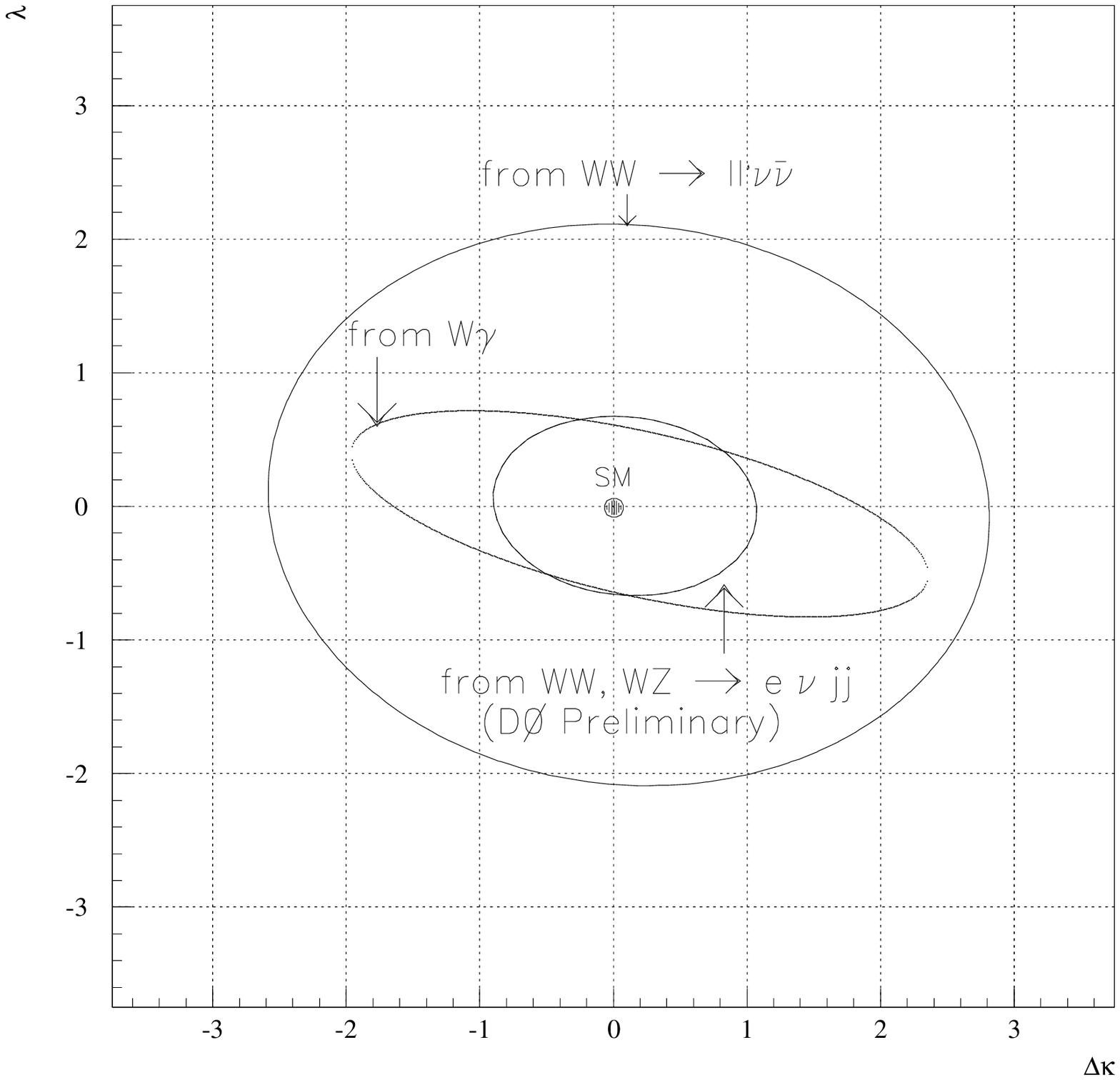,bbllx=0pt,bblly=100pt,bburx=550pt,bbury=700pt,
width=15cm}}
\caption{
Comparison of the limit obtained in this paper with limits
obtained from a measurement of $W\gamma$
production~[4] and a search for
$WW\rightarrow \ell\ell^\prime\nu\bar{\nu}^\prime$~[5].}
\label{compare}
\end{figure}

The absence of an excess of events with high $p_T^W$ excludes large
deviations from the SM couplings.
To set limits on the anomalous coupling parameters, a binned likelihood
fit was performed on the $p_T$ spectrum of the $e\nu$ system,
by calculating the probability for the sum of the background and
the Monte Carlo signal prediction
as a function of anomalous coupling parameters,
to fluctuate to the observed number of events.
The uncertainties in the background estimate, efficiencies, acceptance and
integrated luminosity were convoluted in the likelihood function
with Gaussian distributions.
Figure~\ref{WW:contour} shows the
limit contour at the $95\%$ CL for the CP--conserving
anomalous coupling parameters,
assuming that CP--violating anomalous coupling parameters are zero and that
the $WWZ$ coupling parameters are equal to the $WW\gamma$ coupling
parameters:
$\Delta\kappa\equiv\Delta\kappa_{\gamma}=\Delta\kappa_{Z}$ and
$\lambda\equiv\lambda_{\gamma}=\lambda_{Z}$.
We obtained limits at the $95\%$ CL of
$$-0.89<\Delta\kappa<1.07~~(\lambda=0),\ \ \ \
-0.66<\lambda<0.67~~(\Delta\kappa=0),$$
for $\hat{s}=0$ (i.e. the static limit).
The limits obtained are  within the constraints imposed by the S--matrix
unitarity for $\Lambda=1.5$~TeV.
Figure~\ref{compare} compares the limits obtained in this paper with limits
obtained by D\O\ from a measurement of $W\gamma$
production~\cite{D0:Wgamma} and a search for
$WW\rightarrow \ell\ell^\prime\nu\bar{\nu}^\prime$~\cite{D0:WW}.
The preliminary result obtained from this analysis
gives the most stringent limit on $\Delta\kappa.$

%%Work is in progress to derive limits without the assumption of
%%$\Delta\kappa_{\gamma}=\Delta\kappa_{Z}$ and
%%$\lambda_{\gamma}=\lambda_{Z}$.

\section*{Acknowledgments}
We thank the Fermilab Accelerator, Computing, and Research Divisions, and
the support staffs at the collaborating institutions for their contributions
to the success of this work.   We also acknowledge the support of the
U.S. Department of Energy,
the U.S. National Science Foundation,
the Commissariat \`a L'Energie Atomique in France,
the Ministry for Atomic Energy and the Ministry of Science and
Technology Policy in Russia,
CNPq in Brazil,
the Departments of Atomic Energy and Science and Education in India,
Colciencias in Colombia, CONACyT in Mexico,
the Ministry of Education, Research Foundation and KOSEF in Korea
and the A.P. Sloan Foundation.

\begin{references}

% LIST_OF_VISITOR_ADDRESSES.TEX                            01/03/95
%
\bibitem[*]{beijing}
Visitor from IHEP, Beijing, China.

\bibitem[\ddag]{conicet}
Visitor from CONICET, Argentina.

\bibitem[\S]{buenosaires}
Visitor from Universidad de Buenos Aires, Argentina.

\bibitem[\P]{ecuador}
Visitor from Univ. San Francisco de Quito, Ecuador.

\vskip 0.25cm

\bibitem{DPF}H.~Aihara {\it et al.},
 Anomalous Gauge Boson Interactions: Summary of the DPF Working
 Subgroup on Anomalous Gauge Boson Interactions of the DPF Long
 Range Planning Study,
 LBL-37155, MAD/PH/871, UB-HET-95-01, UdeM-GPP-TH-95-14,
 1995,
 and references therein.

\bibitem{Hagiwara}K.~Hagiwara, J.~Woodside and D.~Zeppenfeld,
 Phys. Rev. D {\bf 41}, 2113 (1990).

\bibitem{Peccei}K.~Hagiwara, R.D.~Peccei, D.~Zeppenfeld, and K.~Hikasa,
 Nucl. Phys. {\bf B282}, 253 (1987).

\bibitem{D0:Wgamma}D\O\ Collaboration, S.~Abachi {\it et al.},
 Measurement of $WW\gamma$ gauge boson couplings in $p\bar{p}$
 Collisions at $\sqrt{s}=1.8~{\rm TeV}$,
 FERMILAB-PUB-1995/101-E,
 to be published in Phys. Rev. Lett.

\bibitem{D0:WW}D\O\ Collaboration, S.~Abachi {\it et al.},
 Search for $W$ boson pair production in $\bar{p}p$
 collisions at $\sqrt{s}=1.8~{\rm TeV}$,
 FERMILAB-PUB-1995/044-E,
 to be published in Phys. Rev. Lett.

\bibitem{D0:Zgamma}D\O\ Collaboration, S.~Abachi {\it et al.},
 Measurement of $ZZ\gamma$ and $Z\gamma\gamma$ couplings in $p\bar{p}$
 collisions at $\sqrt{s}=1.8~{\rm TeV}$,
 FERMILAB-PUB-1995/042-E,
 to be published in Phys. Rev. Lett.

\bibitem{CDF:WW}CDF Collaboration, F.~Abe {\it et al.},
%% Limits on $WWZ$ and $WW\gamma$ couplings from $WW$ and $WZ$
%% production in $\bar{p}p$ collisions at $\sqrt{s}=1.8~{\rm TeV}$,
FERMILAB-PUB-1995/036-E,
submitted to Phys. Rev. Lett.

\bibitem{TOP_PRD}D\O\ Collaboration,
S.~Abachi {\it et al.}, Top Quark Search with the D\O\ 1992--93
 Data Sample, FERMILAB-PUB-1995/020-E,
 submitted to Phys. Rev. D.

\bibitem{VECBOS}W.~Giele, E.~Glover and D.~Kosower, Nucl. Phys. {\bf B403}
 (1993)633.

\bibitem{ISAJET}F.~Paige and S.~Protopopescu, BNL Report BNL38034, 1986
 (unpublished), release V6.49.

\bibitem{GEANT}F.~Carminati {\it et. al}, ``GEANT Users Guide," CERN Program
 Library, December 1991 (unpublished).

\bibitem{Xsection}D\O\ Collaboration, S.~Abachi {\it et. al},
 $W$ and $Z$ Boson Production in $p\bar{p}$ Collisions at $\sqrt{s}=1.8$ TeV,
 FERMILAB-PUB-95/130-E,
 to be published in Phys. Rev. Lett.

\bibitem{PYTHIA}H.-U.~Bengtsson and T.~Sj\"{o}strand,\ Computer\ Physics
 Commun.\ {\bf 46}(1987)43.
\bibitem{MC:Zeppenfeld}D.~Zeppenfeld, private communication.

%%\bibitem{Ohnemus}U.~Baur, T.~Han and J.~Ohnemus, Phys. Rev. D {\bf 51},
%%3381 (1995).



%%\bibitem{MC:Baur}U. Baur, private communication.


%%%\bibitem{UA2}J. Alitti {\it et al.},
%%% Direct measurement of the $W-\gamma$ coupling at the CERN $\bar{p}p$
%%% collider,
%%% \pl {\bf B277,} 194-202 (1992)
%%%\bibitem{CDF:Wgamma}F. Abe {\it et al.},
%%%Measurement of $W$-Photon Couplings in $p\bar{p}$ Collisions at
%%% $\sqrt{s}=1.8~{\rm TeV}$,
%%% \prl {\bf 74} 1936-1940 (1995)
%%%\bibitem{CDF:Zgamma}F. Abe {\it et al.},
%%% Limits on $Z$-Photon Couplings from $p\bar{p}$ Interactions at
%%% $\sqrt{s}=1.8~{\rm TeV}$,
%%% \prl {\bf 74} 1941-1945 (1995)

\end{references}
\end{document}